\documentclass[12pt]{article}

\def\reff#1{(\ref{#1})}                      
\def\beq{\begin{equation}}
\def\eeq#1{\label{#1}\end{equation}}               
\providecommand{\abs}[1]{\lvert#1\rvert}
\usepackage{graphicx,amsmath}
\begin{document}
\title{Casimir Energy of a Relativistic Perfect Fluid Confined to a D-dimensional Hypercube}
\author{Ariel Edery${}^{1,2,}$\thanks{Email: edery@mailaps.org}\\\,\\${}^1$ research partly done at McGill University, Physics Department \\3600 University St., Montreal, Quebec, Canada H3A 2T8\\\,\\${}^2$ currently at AI Solutions Inc., 10001 Derekwood Lane, \\suite 215, Lanham, MD 20706.}
\date{}

\maketitle

\begin{abstract}
Compact formulas are obtained for the Casimir energy of a relativistic perfect fluid confined to a $D$-dimensional hypercube with von Neumann or Dirichlet boundary conditions. The formulas are conveniently expressed as a finite sum of the well-known gamma and Riemann zeta functions. Emphasis is placed on the mathematical technique used to extract the Casimir energy from a $D$-dimensional infinite sum regularized with an exponential cut-off. Numerical calculations show that initially the Dirichlet energy decreases rapidly in magnitude and oscillates in sign, being positive for even $D$ and negative for odd $D$. This oscillating pattern stops abruptly at the critical dimension of $D=36$ after which the energy remains negative and the magnitude increases. We show that numerical calculations performed with 16-digit precision are inaccurate at higher values of $D$.  
\end{abstract}

\newpage

\subsection*{\center{I. Introduction}}

If a system has boundary conditions, the infinite vacuum energy is slightly altered compared to the the free continuum case; this leads to a force on the boundaries called the Casimir force. In 1948, Casimir \cite{Casimir} calculated the attractive force between two conducting plane-parallel plates in vacuum due to the zero-point fluctuations of the electromagnetic field. There has been an enormous amount of theoretical work on the subject since the pioneering work of Casimir (for a general review up to 1997 we refer the reader to \cite{Plunien,Mosta}). The earliest experiment to test Casimir's calculation was carried out by Sparnaay \cite{Sparnaay} in 1958. The results were inconclusive due to large systematic errors and uncontrollable electrostatic forces leading to a $100\%$ uncertainty in the results. In 1997, a landmark experiment \cite{Lamoreaux,Lamoreaux2} using a torsion pendulum improved significantly on previous results. The most recent experiments using atomic force microscopes \cite{Mohideen} and high precision capacitance bridges \cite{Chan} are now in agreement with theoretical calculations to within $1\%$, eliminating any doubt as to the reality of the Casimir force. 

In this work, we calculate the Casimir energy for phonons in a relativistic perfect fluid confined to a $D$-dimensional hypercube using the cut-off method. The Casimir energy of a scalar field in a rectangular cavity with $p$ sides of lengths $a_1,a_2,...a_p$ and $D-p$ sides of characteristic length $L>>a_i$ was calculated in \cite{Wolfram} using the Epstein zeta function regularization scheme. It was shown that Neumann and periodic boundary conditions yield a negative Casimir energy. Determining the sign for Dirichlet boundary conditions turned out to be more complicated and was studied in detail in \cite{Caruso,Li} where Epstein zeta function regularization was again employed. In \cite{Caruso} it was shown that in a rectangular cavity with $p$ sides of equal length $L$ and $D-p$ sides of length $>>L$, the sign of the Dirichlet energy depends on whether $p$ is even or odd. For even values of $p$, the energy is positive when $D$ is less than a critical value $D_c$ and negative when $D$ is above $D_c$. For odd values of $p$ the sign is always negative and no critical dimension exists. It was later shown \cite{Li} that it is possible for the Dirichlet energy to be positive for odd values of $p$ if the sides have unequal lengths. 

One alternative to zeta function regularization is the exponential cut-off. The cut-off method was employed in \cite{Svaiter} to calculate the Casimir energy of scalar fields confined to parallel plates in higher dimensions. In our article, we apply the cut-off method to a perfect fluid confined to a $D$-dimensional hypercube. We develop a mathematical technique that enables us to extract the relevant Casimir term from a $D$-dimensional infinite sum: one that contains the square root of a sum of $D$ squares modified by an exponential cut-off term. This technique makes repeated use of the Euler-Maclaurin integration formula and a series expansion for the infinite sum of modified Bessel functions. For both von Neumann and Dirichlet boundary conditions we obtain convenient formulas for the Casimir energy as a function of $D$. The formulas are expressed as a single sum of $D$ terms containing the  Riemann zeta and gamma functions. Numerical calculations show that the Dirichlet energy exhibits a clear oscillating pattern up to $D=35$: it is positive for even $D$, negative for odd $D$ and its magnitude decreases rapidly. However, this oscillating pattern stops abruptly at the critical dimension of $D=36$; for $D\ge 36$, the sign remains negative and the magnitude increases. In contrast to the Dirichlet energy, the Neumann energy is negative for all values of $D$. 

It is instructive to compare the Casimir calculation of a perfect fluid to that of the open bosonic string. A string embedded in $D$ spatial dimensions supports transverse vibrations in $D-1$ orthogonal directions. The boundary conditions at the two ends of the string, responsible for the Casimir effect, is independent of the dimension $D$. Therefore, the number of dimensions does not complicate the Casimir calculation: the quantity $D-1$ contributes only a multiplicative factor (see \cite{Polchinski} for details). A fluid confined to a $D$-dimensional hypercube supports longitudinal vibrations in $D$ orthogonal directions. In contrast to the string, it has boundary conditions in all $D$ directions leading to a Casimir energy with a non-trivial dependence on $D$. Simply put, for the open bosonic string one needs to calculate a single infinite sum which is multiplied $D-1$ times whereas for the fluid one needs to calculate a $D$-dimensional infinite sum. This reflects the fact that the perfect fluid is described by one scalar field which is a function of $D+1$ spacetime dimensions whereas the string is described by $D+1$ scalar fields each a function of two spacetime dimensions. 

\subsection*{\center{II. The Acoustic Modes in a Relativistic Perfect Fluid}}

A perfect fluid is defined as having at each point a velocity {\bf v} such that an observer moving with this velocity observes the fluid as being isotropic. This occurs when the mean free path between collisions is small compared to the wavelength. In a frame where the fluid is at rest at some particular position and time the energy-momentum tensor $T^{\mu\,\nu}$ has spherical symmetry and is given by \cite{Weinberg} 
\beq
T^{ij} = P \delta^{ij} \,;\quad  T^{i0} = 0 \,; T^{00} = \rho \,
\eeq{energy-mom}
where $\rho$ is defined as the proper energy density and $P$ the pressure.
In the same frame the current four-vector $N^{\mu}$ is given by
\beq
N^{i}=0 \,; N^{0}=n 
\eeq{N}
where $n$ is defined as the particle number density. The motion of the fluid is governed by conservation of energy-momentum and particle number i.e.
\beq
\partial_{\alpha}T^{\alpha\beta} = 0 \,; \quad \partial_{\alpha}N^{\alpha}=0\,.
\eeq{conserve}
Small perturbations from equilibrium ($\rho=\rho_{0}$, $P=P_{0}$ and $n=n_{0}$) lead to sound waves with the following scalar equation
\beq
\dfrac{\partial^{2}\rho(x)}{\partial^{2}\,t} - v^{2} \nabla^{2}\rho(x)= 0
\eeq{sound}
where $\rho(x)$ is a scalar field, $v$ is the speed of the sound waves given by $v=\sqrt{\tfrac{P_0}{\rho_0}}$ and $x= ({\bf x},t)$. Consider the fluid confined to a D-dimensional hypercube with sides of length $L$. The von Neumann (N) and Dirichlet (Di) boundary conditions at $x^{i}=0$ and $x^{i}=L$ are $ \partial^{i}\rho(x)=0$ and $\rho(x)=0$ respectively (where i=1,2,...,D). The solution to the wave equation \reff{sound} for the von Neumann and Dirichlet boundary conditions are respectively,  
\beq
\rho(x)=\sum_{\{n_i\}=0}^\infty
\left(\alpha_{\{n_i\}}^{\dagger} e^{i\omega\,t} + \alpha_{\{n_i\}} \,e^{-i\omega\,t}\, \right)\prod_{i=1}^{D}
\cos(\dfrac{n_i\,\pi\,x^{i}}{L})
+ \,a\,t + \,b 
\eeq{SolNeumann}
and
\beq
\rho(x)=\sum_{\{n_i\}=1}^\infty
(\alpha_{\{n_i\}}^{\dagger} e^{i\omega\,t} + \alpha_{\{n_i\}} \,e^{-i\omega\,t}\,)\prod_{i=1}^{D}
\sin(\dfrac{n_i\,\pi\,x^{i}}{L})
\eeq{SolDirichlet}
where $\omega$ is given by
\beq
\omega = \dfrac{\pi\,v}{L}\,(n_1^2 + n_2^2 + \cdots + n_D^2)^{1/2}
= \pi\,\beta\,(n_1^2 + n_2^2 + \cdots + n_D^2)^{1/2} \,. 
\eeq{omega}
The parameter $\beta\equiv v/L$ is dependent on the physical and geometrical properties of the fluid: the pressure $P_{0}$, the proper density $\rho_{0}$ and the proper length $L$ of the sides of the hypercube. 

\subsection*{\center{III. Quantization and Casimir Energy}}

After imposing equal time commutation relations on the scalar field $\rho(x)$ i.e.
\beq
[\rho({\bf x},t), \dot{\rho}({\bf x}^{\prime},t)]=i\delta^D ({\bf x}-{\bf x}^\prime)
\eeq{commute}
one obtains the well known form for the vacuum energy $E=\tfrac{1}{2}\sum\omega$ (where $\hbar=1$). For the D-dimensional perfect fluid in consideration, $\omega$ is given by \reff{omega} and the vacuum energy in the Neumann(N) and Dirichlet(Di) cases are 
\beq
E= \dfrac{\pi \,\beta}{2}\sum_{\{n_i\}{= 0\,(N)\atop \,= 1 \,(Di)}}^{\infty}(n_1^2 + n_2^2 + \cdots + n_D^2)^{1/2}. 
\eeq{vacuum}
The multiple sum corresponds to the vacuum energy of the fluid with boundary conditions and is divergent due to the high-frequency modes. The vacuum energy with no boundaries i.e. of the continuum, is given by multiple integrals and is also divergent. It is the difference between these two energies that is of interest and leads to the finite quantity we call the Casimir Energy (the energy needed to set up the boundaries starting from the continuum). To extract the relevant constant from the infinite sum \reff{vacuum}, one regularizes the sum to isolate the infinite contribution of the continuum from the finite contribution stemming from the boundary conditions. There are many ways to regularize a sum. In this article we choose an exponential cut-off term $e^{- \,a \,(n_1^2 + n_2^2 + \cdots + n_D^2)^{1/2}}$
where the parameter $a$ is a positive real number. The regularized vacuum energy $E_{Di}$ in the Dirichlet case is then
\begin{eqnarray}
E_{Di} &=&\left(\dfrac{\pi\,\beta}{2}\right)\sum_{n_D=1}^\infty\ldots\sum_{n_1=1}^\infty (n_1^2 + n_2^2 + \cdots + n_D^2)^{1/2}\,\,e^{- \,a \,(n_1^2 + n_2^2 + \cdots + n_D^2)^{1/2}}\nonumber\\
&=&\left(\dfrac{\pi\,\beta}{2}\right)\,\,(-\partial_a) \sum_{n_D=1}^\infty\ldots\sum_{n_1=1}^\infty e^{- \,a \,(n_1^2 + n_2^2 + \cdots + n_D^2)^{1/2}}. 
\label{EN}
\end{eqnarray}
In the Neumann case the sums start at $n_i=0$ instead of $n_i=1$. The regularized vacuum energy \reff{EN} is finite and is a function of the parameter $a$. Our goal is to obtain the leading terms in this sum as $a\rightarrow 0$ and extract the constant Casimir term as a function of the dimension $D$. To accomplish this task we make repeated use of the Euler-Maclaurin integration formula: a formula that relates an infinite sum of a function to its integral i.e.
\beq
\sum_{i=1}^{\infty} f(i) = \int_0^\infty f(x) \,dx - \dfrac{1}{2}\,f(0) - \sum_{p=1}^{\infty}\dfrac{1}{(2p)!}\,B_{2p}\,f^{(2p-1)}(0)
\eeq{Euler}
where $f^{(2p-1)}(0)$ are odd derivatives of $f$ evaluated at zero. There are $D$ sums in \reff{EN} to evaluate and we apply the Euler-Maclaurin formula to each sum except the last one. In Appendix A we show that for the exponential function $f$ in \reff{EN}, the value of $f^{2p-1}(0)$ is always zero except for the last sum. At the last sum, the value of $f^{2p-1}(0)$ can diverge and oscillate between positive and negative infinity (depending on the value of $p$) and we therefore use a different method of calculation. To summarize, we convert $D-1$ sums in \reff{EN} into multiple integrals by repeated application of the Euler-Maclaurin formula and then evaluate separately the last sum. We see from \reff{Euler} that each sum (except the last one) gets replaced by an integral of the function minus half of the function at zero. This can be expressed by a simple and useful prescription 
\beq
\sum \to \int - \dfrac{1}{2} \,.    
\eeq{brevity}
The prescription \reff{brevity} can be applied repeatedly to convert multiple sums to multiple integrals. The case $D=3$ is illustrated below where two of the three sums are replaced by \reff{brevity}:
\begin{multline}
\sum_{n_3=1}^\infty\sum_{n_2=1}^\infty\sum_{n_1=1}^\infty e^{- \,a \,(n_1^2 + n_2^2 + n_3^2)^{1/2}} 
\to \sum_{n_3=1}^\infty \left(\int - \dfrac{1}{2}\right)^2\nonumber\\ 
= \sum_{n_3=1}^\infty\left( \int^2 - \int + \dfrac{1}{4}\right)
=\sum_{n_3=1}^\infty \int_0^\infty e^{- \,a \,(n_1^2 + n_2^2 + n_3^2)^{1/2}} dn_1\,dn_2\\
- \sum_{n_3=1}^\infty \int_0^\infty e^{- \,a \,(n_2^2 + n_3^2)^{1/2}} dn_2
+ \dfrac{1}{4} \sum_{n_3=1}^\infty e^{- \,a \,n_3} \,\,.
\label{cool}
\end{multline}
To evaluate \reff{EN}, we apply $D-1$ times the prescription given in \reff{brevity}. This yields
\begin{eqnarray}
E_{Di} &=& -(\pi\,\beta/2)\,\,\partial_a \sum_{n_D=1}^\infty \left( \,\int - \dfrac{1}{2} \,\right)^{D-1} 
=\pi\,\beta\,(-1)^{D}  \,2^{-D}\,\partial_a \sum_{n_D=1}^\infty \left(\,1 - 2\,\int\,\right)^{D-1} \nonumber \\
&=&  \pi\,\beta\,(-1)^{D} \,2^{-D}\,\sum_{p=0}^{D-1}\dbinom{D-1}{p} (\,-2\,)^p \,\,\partial_a \sum_{n_D=1}^\infty \int^p \nonumber\\
&=&\pi\,\beta\,(-1)^{D} \,2^{-D}\, \sum_{p=0}^{D-1}\dbinom{D-1}{p}\,(\,-2\,)^p \,\partial_a\,I(p,a) 
\label{DSum}
\end{eqnarray}
where $I(p,a)$ is defined by
\beq
I(p,a) \equiv \sum_{n=1}^\infty \int^p = \sum_{n=1}^\infty \int_0^\infty e^{- \,a \,(n^2 + x_1^2 +\ldots + x_p^2)^{1/2}} dx_1\ldots dx_p . 
\eeq{IP}
To determine \reff{DSum} we need to evaluate $\partial_a\,I(p,a)$. 
The p-dimensional integral in $I(p,a)$ can be expressed in terms of the derivative of the Modified Bessel function $K_{\frac{p-1}{2}}(a\,n)$ \cite{Gradshteyn}
\beq
\int_0^\infty e^{- \,a \,(n^2 + x_1^2 +\ldots + x_p^2)^{1/2}} dx_1\ldots dx_p 
=-2^{\frac{1-p}{2}}\,\pi^{\frac{p-1}{2}}\partial_a 
\left(K_{\frac{p-1}{2}}(a\,n) \left(\dfrac{n}{a}\right)^{\frac{p-1}{2}}\right) .
\eeq{xtor}
Using the identity 
\beq
\left(\dfrac{d}{z\,dz}\right)^{m}\{z^{-\nu}\,K_{\nu}(Z)\}=(-1)^m \, Z^{-\nu -m}\, K_{\nu +m}(Z)
\eeq{identity}
with $\nu=0$, $m=\frac{p-1}{2}$, $Z=a\,n$ yields
\beq
(-1)^{\frac{1-p}{2}}\left(\dfrac{d}{a \,da}\right)^{\frac{p-1}{2}} \! \! \! K_{0}(a\,n) =
K_{\frac{p-1}{2}}(a\,n) \left(\dfrac{n}{a}\right)^{\frac{p-1}{2}}\,.
\eeq{besselzero}
By substituting \reff{besselzero} and \reff{xtor} into \reff{IP} one obtains    
\beq
\partial_a I(p,a)= 2^{\frac{1-p}{2}}\pi^{\frac{p-1}{2}} (-1)^{\frac{3-p}{2}}
(\partial_a)^2 \left(\dfrac{d}{a \,da}\right)^{\frac{p-1}{2}} 
\sum_{n=1}^\infty K_{0}(a\,n).
\eeq{IP3}
We are interested in obtaining a series expansion of \reff{IP3} and isolating the relevant constant from the infinite continuum in the limit as $a\to 0$.
We therefore replace the infinite sum of $K_{0}(a\,n)$ by the following series expansion \cite{Gradshteyn}
\begin{eqnarray}
\sum_{n=1}^\infty K_0(a\,n)&=&\dfrac{1}{2}\left(C +\ln(a/4\,\pi)\right) +\dfrac{\pi}{2\,a}\nonumber\\ 
&+& \pi \,\sum_{m=1}^{\infty}\left\{\dfrac{1}{\sqrt{a^2 +4\,m^2\,\pi^2}} - \dfrac{1}{2\,m\,\pi}\right\}. 
\label{sump}
\end{eqnarray}
Consider the terms $\ln(a/4\,\pi)/2$ and $\pi/(2\,a)$ in \reff{sump}. They yield terms proportional to $1/a^{p+1}$ and $1/a^{p+2}$ respectively in the series expansion of \reff{IP3}. These two terms correspond to the infinite continuum as $a\to 0$. The relevant constant related to the Casimir energy stems from the infinite sum in \reff{sump} i.e.
\begin{eqnarray}
\lim_{a \to 0}\!\!\!\!\!\!& &\!\!\!\!\!\!(\partial_a)^2 \left(\dfrac{d}{a \,da}\right)^{\frac{p-1}{2}} 
\sum_{m=1}^{\infty}\dfrac{\pi}{\sqrt{a^2 +4\,m^2\,\pi^2}}\nonumber\\
&=& \lim_{a \to 0} \,\dfrac{(-1)^{\frac{p+1}{2}} \,\Gamma(p+1)}{\Gamma(\frac{p+1}{2})\,2^{\frac{p-1}{2}}}\sum_{m=1}^{\infty}\dfrac{\pi}{(a^2 +4\,m^2\,\pi^2)^{\frac{p+2}{2}}} + O(a) \nonumber\\ 
&=& \dfrac{(-1)^{\frac{p+1}{2}}\,\Gamma(\frac{p+2}{2})}{\pi^{\frac{2p+3}{2}}\,2^{\frac{p+3}{2}}}
\sum_{m=1}^{\infty}\dfrac{1}{m^{p+2}}\nonumber\\
&=&\dfrac{(-1)^{\frac{p+1}{2}}\,\Gamma(\frac{p+2}{2})\,\zeta(p+2)}{\pi^{\frac{2p+3}{2}}\,
2^{\frac{p+3}{2}}}\,.
\label{long}
\end{eqnarray}
Inserting \reff{long} into \reff{IP3} one obtains
\beq
\partial_a I(p,a) = \dfrac{\Gamma(\frac{p+2}{2})\,\zeta(p+2)}{2^{p+1} \,\pi^{\frac{p+4}{2}}\,}.
\eeq{deri}
We finally obtain the Casimir energy for the Dirichlet case 
by substituting \reff{deri} into \reff{DSum} 
\beq
E_{Di}=\beta\,\,2^{(-D-1)}\sum_{p=0}^{D-1}\tbinom{D-1}{p}\,(-1)^{p+D} \,\pi^{\frac{-p-2}{2}}\, 
\,\Gamma(\tfrac{p+2}{2})\,\zeta(p+2)\, .
\eeq{near}
Equation \reff{near} is our final formula for the Casimir energy of a relativistic perfect fluid confined to a hypercube with Dirichlet boundary conditions. It is conveniently expressed as a finite sum of $D$ terms involving the gamma and Riemann zeta functions; this makes it well-suited for numerical calculations. The parameter $\beta$ encompasses the physical and geometrical properties of the relativistic perfect fluid: its proper energy density, pressure and length $L$. It plays the same role for the fluid as the string tension does for the bosonic string; both $\beta$ and the string tension appear in the Casimir energy as dimensionful free parameters.

Having solved the Dirichlet case it is now relatively straightforward to obtain the Neumann case. The regularized vacuum sum in the Neumann case, labelled $E_{N}$, has its sums starting at $n_i=0$ instead of $n_i=1$ i.e.
\beq
E_{N} =\left(\dfrac{\pi\,\beta}{2}\right)\,\,(-\partial_a)  \sum_{n_D=0}^\infty\ldots\sum_{n_1=0}^\infty e^{- \,a \,(n_1^2 + n_2^2 + \cdots + n_D^2)^{1/2}}. 
\eeq{SN}
The above $D$-dimensional sum can be expressed as a series of $k$-dimensional sums that start at $n_i=1$ instead of $n_i=0$ i.e. $k$-dimensional Dirichlet sums. The procedure is as follows: we choose $k$ out of the $D$ sums and let these $k$ sums start at 1 instead of zero (while the remaining $D-k$ variables are not summed and set to zero). One is left with a $k$-dimensional Dirichlet sum $E_{Di}^{(k)}$. There are $\binom{D}{k}$ ways to choose $k$ among $D$ sums so that the Neumann Casimir energy is given by  
\beq
E_{N}=\sum_{k=1}^{D} \tbinom{D}{k} \, E_{Di}^{(k)} 
\eeq{EN2}
where $E_{Di}^{(k)}$ is the $k$-dimensional Dirichlet Casimir energy obtained by replacing 
$D$ by $k$ in \reff{near}. Equations \reff{near} and \reff{EN2} are our final expressions for the Dirichlet and Neumann  Casimir energies respectively. 

In table 1 we quote values of the Dirichlet and Neumann Casimir energies for $D$ up to $6$ calculated using \reff{near} and \reff{EN2}. Note that the Neumann Casimir energy is negative. In Appendix B, we prove that it is negative for all values of $D$. The Neumann energy, plotted in figure 1, has a magnitude which increases with $D$. The Dirichlet case is considerably more complicated. Table 1 shows that the sign of the Dirichlet energy is negative for odd values of $D$, positive for even values of $D$ and that its magnitude decreases rapidly. These features of the Dirichlet energy are valid for low values of $D$ and are plotted in figure 2. The values quoted in table 1 are in agreement with those calculated using the Epstein zeta function in \cite{Wolfram} (values in \cite{Wolfram} are quoted up to $D=5$). It is important that the numerical values agree because the Casimir force should be independent of the regularization scheme employed.

Calculations of the Dirichlet energy at higher values of $D$ reveal that the oscillation of the sign and the rapid decrease in magnitude stops at the critical dimension of $D=36$. The Dirichlet energy decreases by twelve orders of magnitude from $D=1$ to $D=36$. To view a plot over such a large span requires the energy to be scaled. The magnitude of the Dirichlet energy $E$ is less than 1 for the range we consider so that the function $\tfrac{-E}{\abs{E}\,\log(\abs{E})}$ is well-suited for plotting; it preserves the sign and scales the magnitude appropriately. A plot of this function up to $D=110$ is shown in figure 3. The distinctive features of figure 3 are the oscillating pattern which stops abruptly at the critical dimension of $D=36$ and the `plateau' region which emerges immediately afterwards. For $D\ge 36$, the energy remains negative and the magnitude increases, though slowly in the `plateau' region extending to approximately $D=80$.
To obtain accurate values of the Casimir energy at higher values of $D$, numerical calculations must be performed with greater precision than 16-digit precision. We quote in table 2 the Dirichlet energy from $D=10$ to $D=80$ for calculations performed using 16-digit, 24-digit and 50-digit precision. With 16-digit precision, numbers begin to show errors in the first significant digit at $D=42$ and the sign is wrong for the first time at $D=49$ (yields a positive instead of a negatve sign). Note that in 16-digit precision oscillations in the sign resume in the region $D>49$. This is incorrect; higher-precision calculations show that the sign remains negative starting at $D=35$. The plot in figure 3 corresponds to numerical calculations done with 50-digit precision; for our plot up to $D=110$ this is more than enough precision. Note that the 24-digit and 50-digit precision calculations yield identical results for values quoted up to $D=80$ with four significant digits.   

We now summarize our results in light of previous work on the Casimir energy of scalar fields confined to rectangular boundaries. One of our goals was mathematical: to develop a procedure for calculating \reff{EN}, a multi-dimensional infinite sum regularized with an exponential cut-off. By repeated use of the Euler-Maclaurin formula and a series expansion for the infinite sum of the modified Bessel function $K_0(a\,n)$, we were able to isolate the divergent terms and extract the finite Casimir energy. In effect, we reduced \reff{EN} to a finite sum containing only the gamma and Riemann zeta functions i.e. formula \reff{near}. Numerical calculations show that $D=36$ is a critical dimension, being the first even dimension with negative Dirichlet energy. Our work focused on the simple geometry of the hypercube whereas previous work \cite{Wolfram,Caruso,Li} considered the more general rectangular case and employed Epstein zeta function regularization. Results for the rectangular case are expressed in terms of asymptotic formulae. In \cite{Caruso}, the Dirichlet energy for a $D$-dimensional rectangle with $p$ equal sides is conveniently expressed by a single integral with limits running from zero to infinity and integrand containing the elliptic $\theta$ function. In \cite{Wolfram,Caruso}, numerical values are quoted for low values of $D$ for the hypercube case and they are in agreement with our values. However, the finite formula \reff{near} was not derived in \cite{Wolfram,Caruso} for the special case when all sides of the rectangle are equal. 

It is worth noting that the original sum \reff{vacuum} is mathematically a special case of a more general class of multiple sums involving arbitrary exponents i.e.
\beq
M(s;a_1,\ldots,a_D;\alpha_1,\ldots,\alpha_D;c)=\sum_{n_1,...,n_D=1}^\infty (a_1\,n_1^{\alpha_1} + ...+a_D\,n_D^{\alpha_D}+c)^{-s}.
\eeq{general}
Using zeta function regularization, Elizalde \cite{Elizalde} obtained explicit formulae for \reff{general} expressed as an asymptotic expansion containing the Riemann and Hurwitz zeta functions. In contrast to the exponential cut-off method, zeta function regularization does not require the introduction of new terms like exponentials for convergence; one starts with a convergent sum like \reff{general} valid for Re $s >0$ big enough and then one makes an analytical (usually meromorphic) continuation to other values of $s$. For a detailed mathematical treatment of the zeta function regularization theorem and its  applications to the Casimir energy, the reader is referred to \cite{Elizalde2}. 

\subsection*{\center{Acknowledgments}}
I wish to thank Professor Manu Paranjape and the Particle Physics group of the University of Montreal for their invitation to present a talk on this subject at the Montreal Joint High Energy Physics Seminars on March 28, 2002. I also wish to thank Dr Paranjape for his useful comments and suggestions on the manuscript.
I thank the referee for bringing to my attention the work of Emilio Elizalde.
\begin{appendix}
\section{Appendix}
In applying the Euler-Maclaurin formula \reff{Euler} to the sums in \reff{EN}, we show that $f^{2p-1}(0)=0$ (except when applied to the last sum in \reff{EN}). The most general form for the function $f$ is a $k$-dimensional integral  
\beq
f \equiv\int_0^\infty e^{- \,a \,(n_1^2 +\cdots + n_q^2 + x_1^2 +\cdots + x_k^2)^{1/2}} dx_1\ldots dx_k 
= \int_0^\infty G \,\, dx_1\ldots dx_k    
\eeq{f}
where $ G \equiv e^{- \,a \,(n_1^2 +\ldots + n_q^2 + x_1^2 +\ldots + x_k^2)^{1/2}}$.
In $f$ there are $k$ continuous variables $x_1, \ldots, x_k$ which run from zero to infinity and there are $q$ discrete variables $n_1, \ldots, n_q$ which run from one to infinity i.e. $f$ is being summed $q$ times. Our goal is to show that the odd derivatives of $G$ with respect to one of the discrete variables, say $n_1$, evaluated at $n_1 = 0$ is zero i.e. that $G^{2p-1}(0)=0$.  The first derivative of G with respect to $n_1$ is $ G^{\prime} = - a\,n_1\,G \,(n_1^2 +\ldots + n_q^2 + x_1^2 +\ldots + x_k^2)^{- 1/2} = -a \,n_1 G\,H $
where $H \equiv  (n_1^2 +\ldots + n_q^2 + x_1^2 +\ldots + x_k^2)^{- 1/2}$. The derivative of $H$ with respect to $n_1$ is $H^{\prime}= - n_1 \,H^{3}$. Note that $G^{\prime}$ and $H^{\prime}$ are expressed in terms of $G$, $H$, $a$ and $n_1$. Any subsequent derivatives of $G^{\prime}$ will therefore contain terms of the form 
\beq
a^{i}\,n_1^{j}\,H^{l}\,G 
\eeq{Gderiv}
where $i$, $j$ and $l$ are non-negative integers. Every additional derivative of $G$ either increases or decreases $j$ by one. Two consecutive derivatives will therefore produce an $even$ change in $j$. The first derivative of $G$, $G^{\prime}$, has $j=1$ so that an additional even number of derivatives applied to $G^{\prime}$ leads to $j$ being odd and positive. Therefore, odd derivatives of $G$ cannot produce terms with $j = 0$. As long as $G$ and $H^{m}$ do not diverge at $n_1=0$, the terms \reff{Gderiv} are zero at $n_1=0$. Clearly, $G$ does not diverge at $n_1=0$.  $H^{m}$ does not diverge at $n_1=0$ as long as $q \ge 2$ i.e. after $n_1$ is set to zero the denominator in $H^{m}$ is never zero if there exists at least one other discrete variable besides $n_1$. Therefore, if $q \ge 2$, the odd derivatives of $G$ evaluated at $n_1=0$ are zero (and hence the odd derivatives of $f$ evaluated at $n_1=0$ are zero). 

If $n_1$ is the last discrete variable i.e. $q=1$, then $H^{m}$ evaluated at $n_1=0$ diverges at the point where the limits of integration are zero. The last sum is 
therefore calculated using a different method.     

\section{Appendix}
In this Appendix we show that the Neumann  
Casimir energy is negative for all values of $D$. The Neumann Casimir energy is given by \reff{EN2} i.e.
\beq
E_{N}=\sum_{k=1}^{D} \dbinom{D}{k} \,E_{Di}^{(k)} 
\eeq{EN3}
where $E_{Di}^{(k)}$ is the $k$-dimensional Dirichlet Casimir energy obtained by replacing $D$ by $k$ in \reff{near} i.e.
\beq
E_{Di}^{(k)}=\beta\,(-1)^{k} \,2^{-k}\sum_{i=1}^{k}\dbinom{k-1}{i-1} \,\dfrac{(\,-2\,)^{i-1} \, 
\,\Gamma(i)\,\zeta(i+1)}{2^{2i-1}\,\pi^{\frac{i}{2}}\,\Gamma(\frac{i}{2})} \,\,.  
\eeq{near2}
Substituting \reff{near2} into \reff{EN3} one obtains
\begin{eqnarray}
E_{N}&=&\beta\,\sum_{k=1}^{D} \sum_{i=1}^{k}\dbinom{D}{k} \,\dbinom{k-1}{i-1}\,(-1)^{k+i-1} \, \,2^{-i-k}\,\pi^{\frac{-i}{2}} \,\zeta(i+1)\dfrac{
\Gamma(i)}{\Gamma(\frac{i}{2})}\nonumber\\
&=& -\beta\,\sum_{i=1}^{D} \pi^{\frac{-i}{2}} \,\,\zeta(i+1)\dfrac{
\Gamma(i)\,\, 2^{-2i}}{\Gamma(\frac{i}{2})\,(i-1)!} \sum_{k=i}^{D}\dbinom{D}{k} \,(k-1)\ldots(k-i+1)\,\left(\dfrac{-1}{2}\right)^{k-i}
\label{ENR}
\end{eqnarray}
where we used the equality $(-1)^{k+i-1}=-(-1)^{k-i}$. Note the change in the limits of the double sum i.e. $k$ runs now from $i$ to $D$ and $i$ runs now from $1$ to $D$. This change does not affect the double sum because one obtains the same pairs $(k,i)$. To show that $E_{N}$ is negative all we need to show is that the sum over $k$ in \reff{ENR} is positive. The sum over $k$ is
 \begin{eqnarray}
&\,&\sum_{k=i}^{D}\dbinom{D}{k} \,(k-1)\ldots(k-i+1)\,\left(\dfrac{-1}{2}\right)^{k-i}
= \left(\dfrac{d}{dx}\right)^{i-1}\sum_{k=1}^{D} \dbinom{D}{k}\,x^{k-1}|_{x=-1/2}\nonumber\\
&=& \left(\dfrac{d}{dx}\right)^{i-1}\, \left(\dfrac{(x+1)^D -1}{x}\right)|_{x=-1/2}
= \left(\dfrac{d}{dy}\right)^{i-1}\, \left(\dfrac{y^D -1}{y-1}\right)|_{y=1/2}\nonumber\\
&=& \left(\dfrac{d}{dy}\right)^{i-1}\left(y^{D-1}+ y^{D-2} +\cdots +1\right)|_{y=1/2}
\label{ki}
\end{eqnarray}
where $y=x+1$. Clearly derivatives of the polynomial $y^{D-1} + y^{D-2} +\cdots + 1$ evaluated at $y=1/2$ are positive. We therefore have shown that the Neumann Casimir energy is negative for all values of $D$ (assuming the dimension D are positive integers).  
\end{appendix}

\clearpage
\begin{table}\caption{Dirichlet and Neumann Casimir energies in units of $\beta$}
\begin{center}
\begin{tabular}{|c||c|c|c|c|c|c|} \hline 
  &D=1 & D=2 & D=3 & D=4 & D=5 & D=6\\ \hline
$E_{Di}$ & -0.131 & 0.0415 & -0.0157 & 0.00625 &-0.00261 & 0.00112\\ \hline
$E_{N}$ & -0.131 & -0.220 & -0.284 & -0.331 & -0.367 & -0.396\\ \hline
\end{tabular}
\end{center}
\end{table} 
\clearpage
\begin{table}[h]
\begin{center}
\caption{Dirichlet Energy (in units of $\beta$) for Different Precision Calculations}  
\includegraphics[scale=0.70]{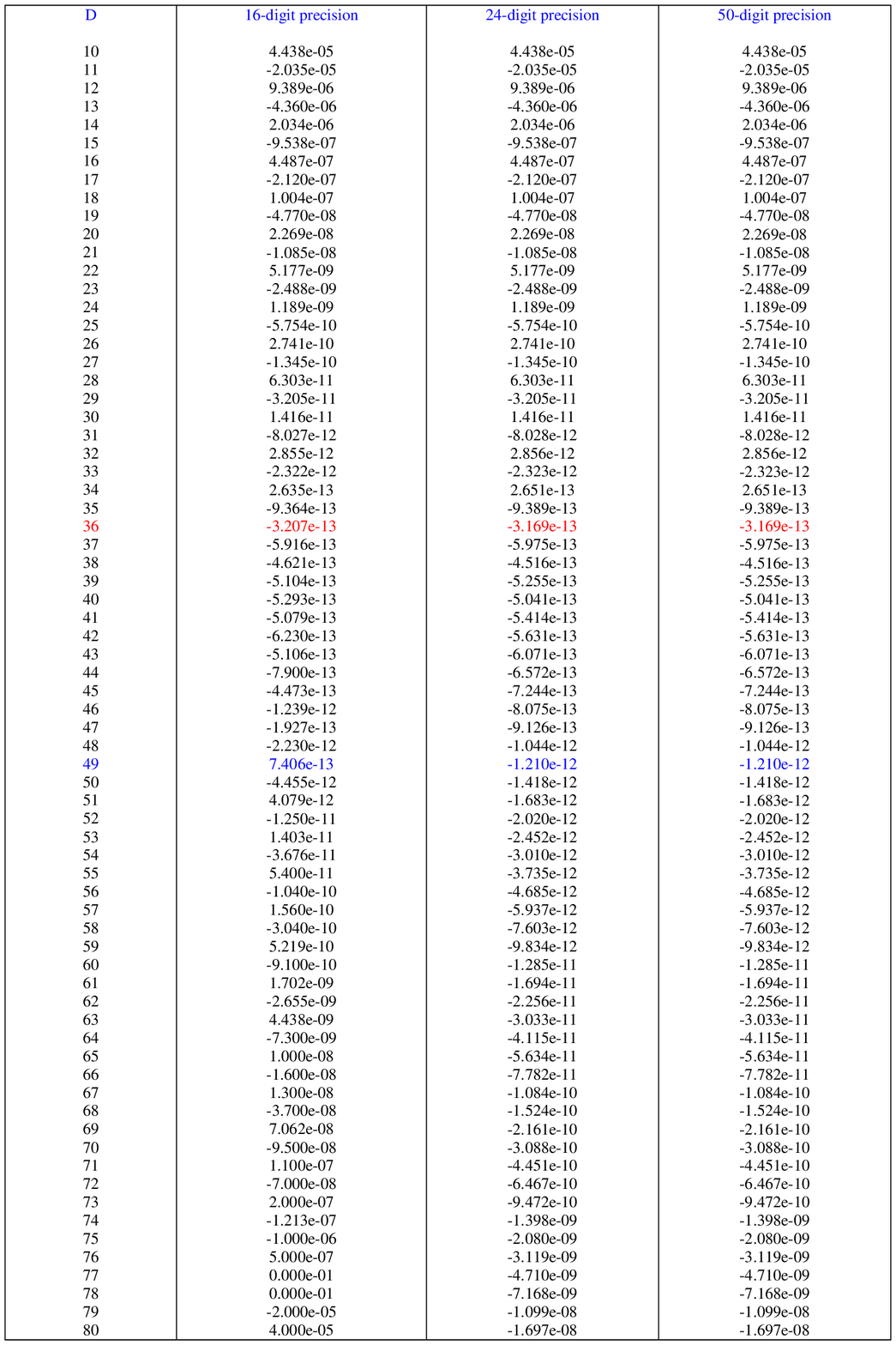}
\end{center}
\end{table}
\clearpage
\begin{figure}
\centering
\caption{Neumann Casimir energy as a function of the dimension D} 
\includegraphics[scale=0.6]{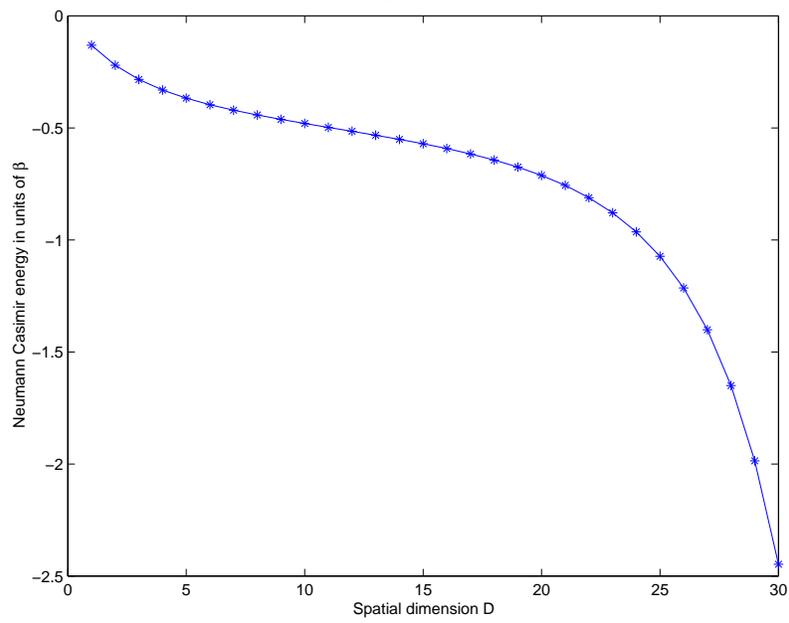}
\end{figure}
\clearpage
\begin{figure}
\centering
\caption{Dirichlet Casimir energy at low values of D} 
\includegraphics[scale=0.6]{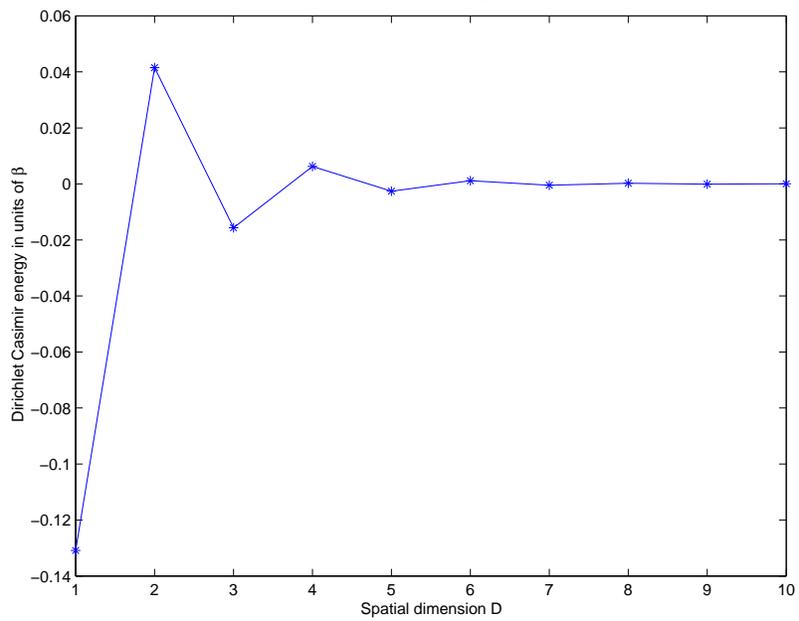}
\end{figure}
\clearpage 
\begin{figure}
\centering
\caption{Scaled Value of the Dirichlet Casimir energy E} 
\includegraphics[scale=0.6]{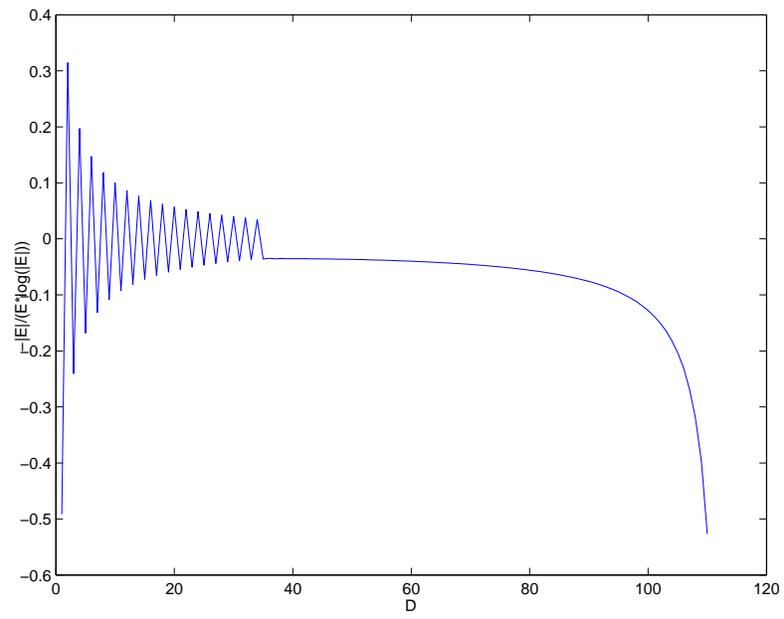}
\end{figure}
\clearpage
\listoffigures
\listoftables

\end{document}